\documentclass{jfm}

\usepackage{graphicx, amsmath} 
\usepackage[usenames]{color}
\usepackage{marvosym}
\usepackage{amssymb}

\let\oldepsilon\epsilon
\let\epsilon\varepsilon
\let\varepsilon\oldepsilon

\title{A two-phase flow model of sediment transport: transition from bedload to suspended load} 
\author[F. Chiodi, P. Claudin and B. Andreotti]
{F\ls I\ls L\ls I\ls P\ls P\ls O\ns  C\ls H\ls I\ls O\ls D\ls I,\ns  P\ls H\ls I\ls L\ls I\ls P\ls P\ls E \ns C\ls L\ls A\ls U\ls D\ls I\ls N, \ns  \and \ns B\ls R\ls U\ls N\ls O\ns A\ls N\ls D\ls R\ls E\ls O\ls T\ls T\ls I} 
\affiliation{
Laboratoire de Physique et M\'ecanique des Milieux H\'et\'erog\`enes,\\
(PMMH UMR 7636 ESPCI  - CNRS - Univ. Paris Diderot - Univ. P. M. Curie)\\
10 rue Vauquelin, 75005 Paris, France.}
\pubyear{???}
\volume{???}
\pagerange{??--??}
\date{\today}
\pdfoutput=1

\begin{document}
\maketitle

\begin{abstract}
The transport of dense particles by a turbulent flow depends on two dimensionless numbers. Depending on the ratio of the shear velocity of the flow to the settling velocity of the particles (or the Rouse number), sediment transport takes place in a thin layer localized at the surface of the sediment bed (bedload) or over the whole water depth (suspended load). Moreover, depending on the sedimentation Reynolds number, the bedload layer is embedded in the viscous sublayer or is larger. We propose here a two-phase flow model able to describe both viscous and turbulent shear flows. Particle migration is described as resulting from normal stresses, but is limited by turbulent mixing and shear-induced diffusion of particles. Using this framework, we theoretically investigate the transition between bedload and suspended load.
\end{abstract}

\section{Introduction}
When a sedimentary bed is sheared by a water flow of sufficient strength, the particles are entrained into motion. In a homogeneous and steady situation, the fluid flow can be characterized by a unique quantity: the shear velocity $u_*$. The flux of sediments transported by the flow is an increasing function of $u_*$, for which numerous transport laws have been proposed, for both turbulent flows (Meyer-Peter \& M\"uller, 1948; Einstein, 1950; Bagnold, 1956; Yalin, 1963; van Rijn, 1984; Ribberink, 1998; Camemen \& Larson, 2005; Wong \& Parker, 2006; Lajeunesse et al., 2010) and laminar flows (Charru \& Mouilleron-Arnould, 2002; Cheng, 2004; Charru et al. 2004, Charru \& Hinch 2006). Despite a wide literature, some fundamental aspects of sediment transport are still only partially understood. For instance, the dynamical mechanisms limiting transport, in particular the role of the bed disorder (Charru, 2006) and turbulent fluctuations (Marchioli et al., 2006; Le Louvetel-Poilly et al., 2009), remain a matter of discussion. Also, derivations of transport laws have a strong empirical or semi-empirical basis, thus lacking more physics-related inputs. Here we investigate the properties of sediment transport using a two-phase continuum description. In particular, we examine the transition from bedload to suspension when the shear velocity is increased.

A classical reference work for two-phase continuum models describing particle-laden flows, is the formulation of Anderson \& Jackson (1967), later revisited by Jackson (1997). With this type of description, Ouriemi et al. (2009) and Aussillous et al. (2013) have addressed the case of bedload transport in laminar sheared flows. Similarly, Revil-Baudard \& Chauchat (2012) have modelled granular sheet flows in the turbulent regime. In their description, these authors decompose the domain into two layers. An upper `fluid layer' where only the fluid phase momentum equation is solved, and a lower `sediment bed layer' where they apply a two-phase description. A layered structure of collisional sheet flows with turbulent suspension has also been proposed by Berzi (2011, 2013), based on kinetic theory of granular gases. Inspired by these works, we propose in this paper a general model which is able to describe sediment transport both in the laminar viscous regime and in the turbulent regime, and which does not require the domain to be split into several layers.

In the literature, considerable emphasis has been placed on the process of averaging the equations of motion and defining stresses (Jackson 2000). However, an important issue beyond the averaging problem is the choice of closures consistent with observations. Here, the fluid phase is described by a Reynolds-dependent mixing length that has been recently proposed in direct numerical simulations of sediment transport (Dur\'an et al., 2012). In the same spirit as Ouriemi et al. (2009), Revil-Baudard \& Chauchat (2013) and Aussillous et al. (2013), the granular phase is described by a constitutive relation based on recent developments made in dense suspension and granular flows (GDR MIDI, 2004; Cassar et al., 2005; Jop et al., 2006; Forterre \& Pouliquen, 2008; Boyer et al., 2011; Andreotti et al. 2012; Trulsson et al. 2012; Lerner et al. 2012). Our model then takes into account both fluid and particle velocity fluctuations.

This paper is constructed as follows. In section~2, we describe our two-phase flow model, putting emphasis on the novelties we propose with respect to the previous works, namely the introduction of a Reynolds stress and particle diffusion induced both by turbulent fluctuations and by the motion of the particles themselves. Section~3 is devoted to the study of homogeneous and steady transport, simplifying the equations under the assumption of a quasi-parallel flow. The results of the model integration are presented in section~4. We discuss the evolution of the velocity, stress and concentration profiles when the shear velocity is increased, as well as the transport law. We end the paper with a summary and draw a few perspectives. 

\section{The two-phase model}
In this paper, we adopt a continuum description of a fluid-particle system. This approximation is valid whenever the various quantities of interest vary slowly at the scale of grains. On the contrary, such a model cannot accurately describe phenomena that involve particle-scale processes, like the sediment transport threshold (see below).

\subsection{Continuity equation}
We define the particle volume fraction $\phi$ so that $1-\phi$ is the fraction of space occupied by the fluid. We respectively denote by $u_i^p$ and $u_i^f$, the particle and fluid Eulerian velocities. We assume incompressibility for both the granular and the fluid phases: the densities $\rho_p$ and $\rho_f$ are considered as constant. The continuity equation then reads:
\begin{eqnarray}
\frac{\partial \phi}{\partial t}+ \frac{\partial u_i^p \phi}{\partial x_i} &=& -\frac{\partial j_i}{\partial x_i} \;,
\label{eq:massp}\\
\frac{\partial (1- \phi)}{\partial t}+ \frac{\partial u_i^f (1-\phi)}{\partial x_i}&=&\frac{\partial j_i}{\partial x_i} \;,
\label{eq:massf}
\end{eqnarray}
where $j_i$ is a particle flux resulting from Reynolds averaging (see Appendix) and equal to the mean product of velocity and volume fraction fluctuations. We assume here that these correlated fluctuations lead to an effective diffusion of particles, quantified by a coefficient $D$:
\begin{equation}
j_i=-D\frac{\partial \phi}{\partial x_i}
\end{equation}
We will propose later on a closure for the diffusion coefficient $D$. Here, it is important to underline that two types of velocity fluctuations will be taken into account: (i) those induced by the shearing of the grains, which are associated with the large non-affine displacements of particles in the dense limit (see details below) and with hydrodynamic interactions in the dilute limit (Eckstein et al., 1977; Leighton \& Acrivos,1987, Nott \& Brady, 1994; Foss \& Brady, 2000) and (ii) those induced by turbulent velocity fluctuations. The introduction of this term corrects a major flaw of previously proposed two-phase models. Whenever particle-borne stresses tend to create a migration of particles, diffusion counteracts and tends to homogenize $\phi$. The formulation chosen here assumes that the system composed of the particles and the fluid is globally incompressible and does not diffuse: adding (\ref{eq:massp}) and (\ref{eq:massf}), one simply gets a relation between the Euler velocities for the two phases:
\begin{equation}
 \frac{\partial [u_i^p \phi +  u_i^f (1-\phi)]}{\partial x_i} =0
\end{equation}
%

\subsection{Equations of motion}
Following Jackson (2000), the equations of motion are simply written as two Eulerian equations expressing the conservation of momentum for each phase:
\begin{eqnarray}
\rho_p  \left( \frac{\partial \phi u_i^p}{\partial t} + \frac{\partial \phi  u_j^p u_i^p}{\partial x_j} \right)&=&\rho_p \phi {g_i}+
\frac{\partial \sigma_{ij}^p}{\partial x_j} + f_i\;,
\label{QMsolide}\\
\rho_f  \left( \frac{\partial (1-\phi) u_i^f}{\partial t} + \frac{\partial (1-\phi) u_j^f  u_i^f}{\partial x_j} \right)&=&\rho_f (1-\phi) {g_i}+
\frac{\partial \sigma_{ij}^f}{\partial x_j} - f_i\;.
\label{QMfluid}
\end{eqnarray}
The force density $f_i$ couples the two equations and represents the average resultant force exerted by the fluid on the particles. Here, $g_i$ is the gravity acceleration and $\sigma_{ij}^p$ and $\sigma_{ij}^f$ are respectively the stresses exerted on the particles and on the fluid. The stresses are additive: the total stress exerted on the effective medium (grains and fluid) is given by $\sigma_{ij}^{p}+\sigma_{ij}^{f}$ and $f_i$ becomes an internal force for this system.

\subsection{Force exerted by one phase on the other}
The standard hypothesis, introduced by Jackson (2000), is to split the force density $\mathbf{f}$ exerted by the liquid on the solid into a component due to the Archimedes effect --~the resultant force exerted on the contour of the solid, replacing the latter by liquid~-- and a drag force due to the relative velocity of the two phases. We write this force in the form
\begin{equation}
\label{forceinteraction}
f_i= \phi \frac{\partial \sigma_{ij}^f}{\partial x_j} + \frac{3}{4} \phi C_d(\mathcal{R}) \rho_f \, \frac{\left|\mathbf{u}^r\right| \, {u}_i^r}{d}\;,
\end{equation}
where $\mathbf{u}^r=\mathbf{u}^f -\mathbf{u}^p$ is the relative velocity and $d$ is the grain size.   The drag coefficient $C_d$ is a function of the grain-based Reynolds number $\mathcal{R}$ defined by
\begin{equation}
\mathcal{R}=\frac{\left|\mathbf{u}^r\right|\;d}{\nu}
\label{Reynolds}
\end{equation}
using the fluid kinematic viscosity $\nu$. We write the drag coefficient in the convenient phenomenological form
\begin{equation}
C_d = \left(C_\infty^{1/2}+s \mathcal{R}^{-1/2} \right)^2
\label{dragcoeff}
\end{equation}
to capture the transition from viscous to inertial drag (Ferguson \& Church, 2004). Both $C_\infty$ and $s$ \emph{a priori} depend on $\phi$. However, for the sake of simplicity, we will neglect this dependence and consider that $s$ is on the order of $\simeq \sqrt{24} \simeq 5$, as in the dilute limit. Similarly, we will take a constant asymptotic drag coefficient $C_\infty$ equal to $1$. 

After inserting the expression of $f_i$, the equations of motion are modifed to
\begin{eqnarray}
\rho_p  \left( \frac{\partial \phi u_i^p}{\partial t} + \frac{\partial \phi  u_j^p u_i^p}{\partial x_j} \right)&=&\rho_p \phi {g_i}+\frac{\partial \sigma_{ij}^p}{\partial x_j} +\phi \frac{\partial \sigma_{ij}^f}{\partial x_j} + \frac{3}{4} \phi \, C_d(\mathcal{R}) \rho_f \frac{\left|\mathbf{u}^r\right| \, {u}_i^r}{d},\nonumber\\
\label{QMsolide1oli}
\rho_f  \left( \frac{\partial (1-\phi) u_i^f}{\partial t} + \frac{\partial (1-\phi) u_j^f  u_i^f}{\partial x_j} \right)&=&\rho_f (1-\phi) {g_i}+(1-\phi) \frac{\partial \sigma_{ij}^f}{\partial x_j} - \frac{3}{4} \phi \, C_d(\mathcal{R}) \rho_f \frac{\left|\mathbf{u}^r\right| \, {u}_i^r}{d}. \nonumber \\
\label{QMfluid1oli}
\end{eqnarray}
%

\subsection{Fluid constitutive relation}
In order to close the equations, one needs to express the stress tensors. For the fluid, we choose a local isotropic constitutive relation which allows us to recover the well-known regimes:
\begin{equation}
\label{sigfluide}
\sigma^{f}_{ij}= -P^f \delta_{ij}+ \eta_{\rm eff} \; {\dot \gamma}^f_{\rm ij}\;+\rho_f (u_i^f j_j+u_j^f j_i), \quad {\rm with} \quad {\dot \gamma}^f_{\rm ij}=\frac{\partial u_i^f}{\partial x_j}+\frac{\partial u_j^f}{\partial x_i}
\end{equation}
The last term is the flux of momentum associated with particle diffusion (see Appendix).
The effective viscosity $\eta_{\rm eff}$ takes into account both the molecular viscosity and the mixing of momentum induced by turbulent fluctuations. The Reynolds stress, which results from correlated velocity fluctuations (see Appendix), is modelled using a Prandtl mixing length closure, which works well for turbulent shear flows:
\begin{equation}
\eta_{\rm eff} =\rho_f\left(\nu+\ell^2 |\dot \gamma^f|\right)
\end{equation}
where $\ell$ is the mixing length and $|\dot \gamma^f|$ is the typical mixing rate.

This formulation neglects the influence of particles on the fluid effective rheology. This equation can be easily generalized to include a multiplicative factor $(1+5 \phi/2)$ in front of the viscosity $\nu$. The Einstein viscosity is then recovered in the limit of dilute suspensions. Even close to the jamming transition, the effective fluid viscosity remains finite; the presence of particles only adds new solid boundary conditions and therefore reduces the fraction of volume where shear is possible. The dominant effect actually results from particle interactions, as discussed below. We have checked that the Einstein correction can then be safely ignored for the description of sediment transport, i.e. that this corrective factor has a negligible effect on the results.  

In the fully developed turbulent regime, far from the sand bed, the mixing length $\ell$ is proportional to the distance $z$ to the bed. Conversely, in the viscous regime, below some Reynolds number $R_t$, there is no velocity fluctuation so that $\ell$ must vanish. A common phenomenological approach  (van Driest's model for instance, see van Driest (1956) or Pope (2000)) is to express the turbulent mixing length as a function of the Reynolds number and $z$. However, this imposes the definition of an interface between the static and mobile zones, below which $\ell$ must vanish. To avoid the need for such an arbitrary definition, we propose instead a differential equation
\begin{equation}
\label{eq_ell}
\frac{d\ell}{dz}= \kappa\left[1-\exp\left(-\sqrt{\frac{\ell u^f}{R_t\nu}}\right)\right]
\end{equation}
where $\kappa\simeq 0.4$ is the von K\'arm\'an constant and the dimensionless parameter $R_t \simeq 7$ is determined to recover the experimental `law of the wall'. The ratio $u^f \ell/\nu$ is the local Reynolds number based on the mixing length. It should be noted that a function other than the exponential can in principle be used, provided that it has the same behaviour in $0$ and $-\infty$, although the present choice provides a quantitative agreement with standard data (Dur\'an et al., 2012). This formulation allows us to define $\ell$ both inside and above the static granular bed.

The mixing length also depends on convective effects. In a first-order closure, this can be encoded as a dependence of $\kappa$ on the Richardson number
\begin{equation}
{\rm Ri} = \frac{N^2}{|\dot \gamma|^2},\quad{\rm with}\quad N^2 = \frac{ (\rho_f-\rho_p)g}{\phi \rho_p + (1-\phi)\rho_f} \, \frac{d\phi}{dz} \, .
\end{equation}
Whenever $N^2>0$, the flow is stably stratified. Here, $N$ is called the Brunt-V\"ais\"al\"a frequency. For the sake of simplicity, we have omitted this dependence on ${\rm Ri}$ but we acknowledge that it could lead to major effects in the context of particle and droplet transport in the convective boundary layer and in the stably stratified upper layer.

Finally, a term must be, in principle, added to Eq.~\ref{sigfluide} to take into account the third-order correlation between velocity and volume fraction fluctuations on the right hand side of Eq.\,\ref{eqappend}. Such a term would correspond to a transport of momentum resulting from a gradient of $\phi$ (independently of the convection effect), but, once again, we wish here to study this two-phase model in its simplest version.

\subsection{Granular constitutive relation}
For the granular phase, we first decompose the stress tensor into pressure $P^p$ and  deviatoric stress $\mathbf{\tilde \sigma^{p}}$:
\begin{equation}
\label{mudeIviscoli1}
\sigma^{p}_{ij}=- P^p\,\delta_{ij}\;+\tilde \sigma^{p}_{ij}\;-\rho_p (u_i^p j_j+u_j^p j_i)
\end{equation}
The last term is the flux of momentum associated with particle diffusion (see Appendix). We focus here on a first-order closure where the strain rate tensor and the volume fraction are the only state variables. We therefore assume that the granular temperature, defined as the variance of velocity fluctuations, is not an independent field but relaxes over a very short time scale to a value determined by the strain rate and the volume fraction. This approximation is excellent in the dense regime $\phi \to \phi_c$, where numerical simulations show no dependence of the rheology on the grain restitution coefficient. One expects it to be less and less accurate in the dilute limit $\phi \to 0$ and when the restitution coefficient is close enough to $1$. As in Ouriemi et al. (2009), we write the constitutive relation as a simple friction law:
\begin{equation}
P^p=\frac{|\mathbf{\tilde \sigma^{p}}|}{\mu(\phi)}
\label{FrictionLaw}
\end{equation}
It should be noted that in Ouriemi et al. (2009) the friction law was expressed in terms of the inertial number $I = \dot \gamma d / \sqrt{P^p/\rho_p}$ (GDR MiDi, 2004). This was made possible thanks to the assumption that the volume fraction in the transport layer was almost a constant, equal to $\phi_c$. In general, the pressure is not a state variable when the variations of $\phi$ are taken into account self-consistently. We therefore express here the stress tensor (including pressure) as a function of the true state variables: $\phi$ and the strain rate tensor. As $I$ is no longer a state variable, a friction law $\mu(\phi)$ becomes the only possible choice. The equality (\ref{FrictionLaw}) becomes an inequality, i.e. a Coulomb failure criterion, for a static bed, for which $|\dot \gamma^p|=0$. The friction coefficient $\mu$ decreases with the volume fraction $\phi$. Here we use a linear expansion around jamming ($\phi=\phi_c$)
\begin{equation}
\label{mudeIviscoli2}
\mu(\phi) \simeq \mu_c+\mu'_c \left(\phi_c-\phi\right),
\end{equation}
with $\mu_c=0.5$ and $\mu'_c = 0.1$.
 
Following recent experimental and numerical results, the granular viscosity is written as:
\begin{equation}
\label{mudeIviscoli}
\tilde \sigma^{p}_{ij}=\psi(\phi) \;\left(\rho_f \nu+\alpha \rho_p d^2 |{\dot \gamma}^p| \right) {\dot \gamma}^p_{\rm ij}, \quad {\rm with} \quad {\dot \gamma}^p_{\rm ij}=\frac{\partial u_i^p}{\partial x_j}+\frac{\partial u_j^p}{\partial x_i}.
\end{equation}
Here, $\alpha$ is the inverse of the Stokes number at which the transition from viscous to inertial suspension takes place.  Trulsson et al. (2012) have found a value of $\alpha \simeq 0.6$ in a 2D simulation. $\psi(\phi)$ is a function that diverges at the critical volume fraction $\phi_c$ as
\begin{equation}
\label{psidephi}
\psi(\phi) = \upsilon \frac{\phi^2}{\left(\phi_c-\phi\right)^{2}},
\end{equation}
where $\upsilon$ is a numerical constant. The experimental values are approximately $\phi_c=0.615$ and $\upsilon=2.1$ (Ovarlez et al., 2006; Bonnoit et al., 2010). It should be noted, however, that the results presented here have been computed for $\alpha=1$, $\phi_c=0.64$ and $\upsilon=1$.

\begin{figure}
\centerline{\includegraphics{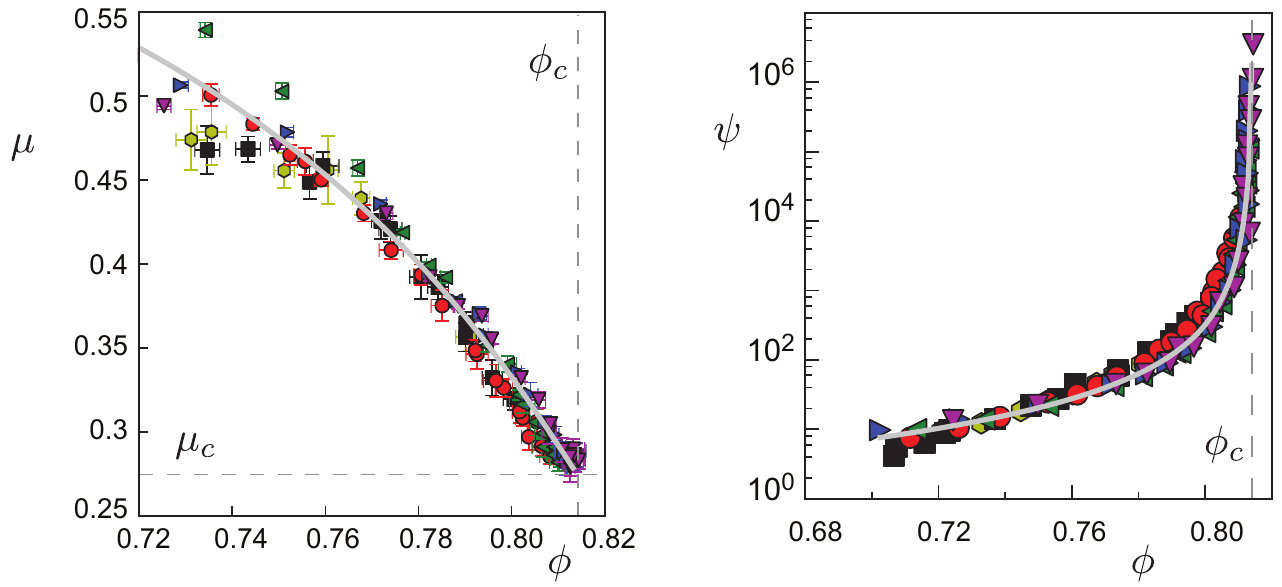}}
\caption{Calibration of the functions $\mu(\phi)$ and $\psi(\phi)$ in a 2D numerical simulation taking into account viscous drag, inertia, contact force, solid friction and lubrication forces (Trulsson et al., 2012). The different colors correspond to different values of the Stokes number $ \rho_p d^2 |{\dot \gamma}^p|/ \rho_f \nu$, which characterizes the relative amplitude of the two terms in Eq.~\ref{mudeIviscoli}.}
\label{Fig1}
\end{figure}

In the dilute limit, the dynamics of suspension is determined by long-range hydrodynamic interactions (Eckstein et al., 1977; Leighton \& Acrivos,1987, Nott \& Brady, 1994; Foss \& Brady, 2000). However, close to jamming, the mechanical properties of dense suspensions are related to the contact network geometry (Wyart et al. 2005; van Hecke, 2010; Lerner et al., 2012). Due to steric effects, the grains in a linear shear flow do not follow the average flow, but move cooperatively following "floppy" modes. This deviation of the actual motion with respect to the mean motion is a type of fluctuation called non-affine motion of particles. To move by a distance as the crow flies equal to its diameter $d$, a grain makes a random-like motion whose average length diverges as $\sim d (\phi_c-\phi)^{-1}$. The  statistical properties of grain trajectories and in particular their cooperative non-affine motions are mostly controlled by the volume fraction $\phi$, whatever the nature of the dissipative mechanisms (Andreotti et al., 2012).  The enhanced viscosity close to jamming is directly related to the amplitude of non-affine motion, leading to a stress tensor diverging as $\psi(\phi)\sim (\phi_c-\phi)^{-2}$ (Boyer et al., 2011; Andreotti et al., 2012; Trulsson et al., 2012; Lerner et al., 2012). Numerical simulations have furthermore suggested that the function $\mu(\phi)$ does not depend on the microscopic interparticle friction coefficient nor on the flow regime (overdamped or inertial), see Fig.~\ref{Fig1}. However, the critical volume fraction $\phi_c$ does depend on this microscopic interparticle friction coefficient. This constitutes the most important consequence of having lubricated quasi-contacts rather than true frictional ones.

Here again, we argue due to the simplicity of the asymptotic expression of $\psi(\phi)$ one should use it, even in the dilute limit. The model could be easily generalized to a more complex choice for $\psi(\phi)$. We have checked that the results presented here do not depend much on such refinements.

\subsection{Diffusion}
Apart from convective transport associated with the mean flow, velocity fluctuations lead to a mixing of particles and tend to homogenize the volume fraction $\phi$. We model these effects by two diffusive terms. The first is associated to turbulent fluctuations. Turbulent diffusivity is usually found to scale with the turbulent viscosity and is thus proportional to $\ell^2 |\dot \gamma^f|$. The second source of diffusion, dominant in the dense regime, originates from the non-affine motion of particles. The particle-induced diffusivity therefore scales as $d^2  |\dot \gamma^p| \psi(\phi)$. A similar scaling law has also been proposed to describe shear-induced diffusion in relatively dilute suspensions (Eckstein et al., 1977; Leighton \& Acrivos,1987). In that case, the equivalent of the function $\psi(\phi)$, typically increasing like $\phi^2$ up to $\phi \simeq 30 \%$, represents the interactions between particles mediated by hydrodynamics. This diffusivity has been measured from simulations based on Stokesian dynamics (e.g., Nott \& Brady, 1994; Foss \& Brady, 2000). Adding the two contributions, the diffusion coefficient reads
\begin{equation}
D= \frac{\ell^2 |\dot \gamma^f|}{{\rm Sc}}+\frac{d^2  |\dot \gamma^p| \psi(\phi)}{S_\phi}
\label{eq:Sc}
\end{equation}
The constant  ${\rm Sc}$ is called the turbulent Schmidt number and lies in the range $0.5$--$1$ (Coleman 1970; Celik \& Rodi 1988; Nielsen 1992). For the results presented here, we have chosen ${\rm Sc}=1$. We have introduced here a second phenomenological constant $S_\phi$, which we take equal to $1$ as well.

\section{Homogeneous steady transport}
\subsection{Dimensional analysis}
We now apply this two-phase flow model to the description of sediment transport. In this article, we limit ourselves to the analysis of saturated transport over a flat bed, for which both phases are homogeneous along the flow ($x$) direction (but not along $z$) and in a steady state. It is worth emphasizing that vertical velocities do not vanish. The total vertical mass flux can still vanish, due to particle diffusion along the vertical direction. As discussed below, the condition of stationarity is ensured when the erosion rate of the bed is exactly balanced by a deposition rate induced by the settling of the particles.

In the following, we will keep the equations dimensional. It is still interesting to perform the dimensional analysis of the problem. We will consider that the characteristic length is set by the grain diameter $d$, that the characteristic density is given by $\rho_f$ and that $(\rho_p-\rho_f) gd$ provides the characteristic stress.  Apart from the constants of the model, there are two control parameters and therefore two dimensionless numbers. The first, called the Shields number, is the rescaled shear stress:
\begin{equation}
\Theta=\frac{\rho_f u_*^2}{(\rho_p-\rho_f) gd} \, .
\end{equation}
It compares the fluid-borne shear stress with the buoyancy-free gravity.  The second parameter is based on the kinematic viscosity $\nu$ of the fluid. Combining this viscosity with gravity, one builds the viscous diameter:
\begin{equation}
d_\nu=\left(\frac{\rho_p}{\rho_f}-1\right)^{-1/3}~\nu^{2/3}~g^{-1/3},
\end{equation}
which corresponds to the grain size for which inertial, gravity and viscous effects are of the same order of magnitude. The dimensionless number is then the ratio $d/d_\nu$. Equivalently, the Galileo number is defined as $(d/d_\nu)^3$; the Reynolds number in Stokes sedimentation is $\frac{1}{18}(d/d_\nu)^3$. Other dimensionless numbers, like the particle Reynolds number in the flow are obtained by combining the Shields number $\Theta$ and the dimensionless diameter  $d/d_\nu$. 

\subsection{Quasi-parallel flow assumption}
We consider a flat sand bed homogeneous along the $x$-axis. The continuity equation then reads

\begin{equation}
\label{eq_phi}
\phi u^p_z=(\phi-1) u^f_z=-j_z=D\frac{d\phi}{dz}\,.
\end{equation}
It expresses the balance between the convective flux associated with the downward particle motion and the turbulent diffusive flux. As they are related by the continuity equation, the vertical velocities can be expressed as functions of their relative velocity $u^r_z$: $u^f_z=\phi u^r_z$ and $u^p_z=(\phi-1) u^r_z$. The diffusive flux then reads $j_z=\phi (1-\phi) u^r_z$. We consider the asymptotic limit where horizontal velocities are much larger than vertical ones. 

\begin{figure}
\centerline{\includegraphics{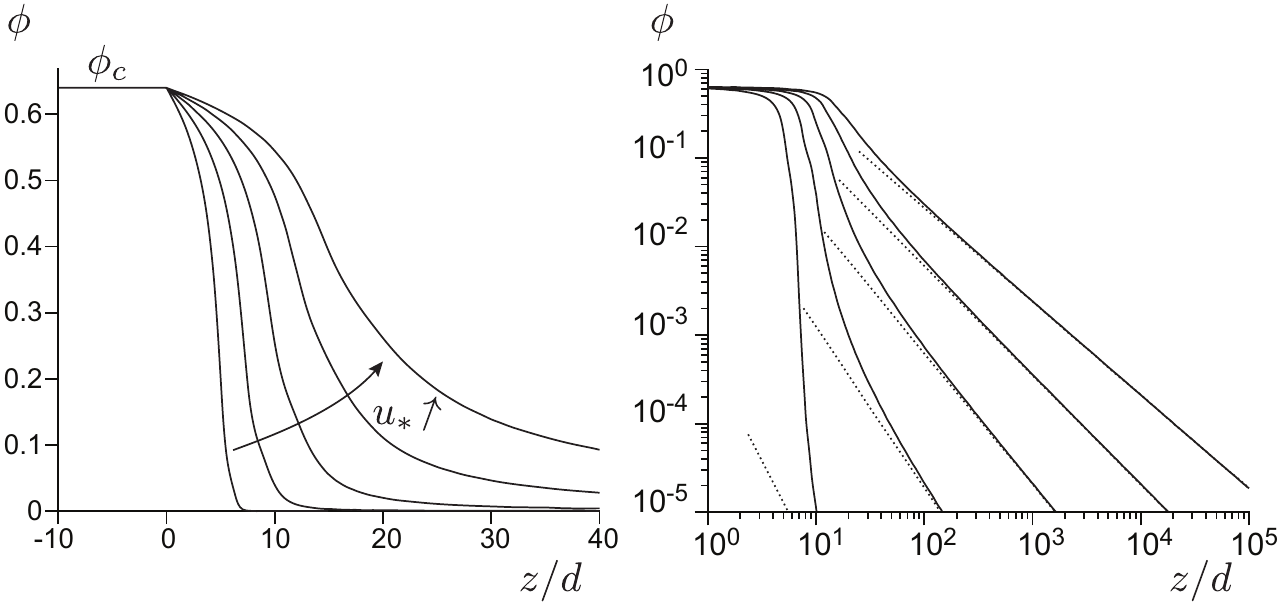}}
\caption{Vertical profiles $\phi(z)$ of the volume fraction in lin-lin (left) and log-log (right) representation, for five values of the shear velocity: $u_*/\sqrt{\left(\frac{\rho_p}{\rho_f}-1\right)gd}=0.3, 0.4, 0.5, 0.6, 0.7$. The dotted line correspond to the power-law obtained in the dilute asymptotic limit (Eq.~\ref{phipowerlaw}).}
\label{Fig2}
\end{figure}

Making use of the homogeneity along $x$, the vertical equations of motion simplify to
\begin{eqnarray}
\rho_p  \frac{d  \phi  \left(u_z^p\right)^2}{d  z} &=&-(\rho_p-\rho_f) \phi g +\frac{d  \sigma_{zz}^p}{d  z} +\phi \frac{d  \sigma_{zz}^f}{d z} + \frac{3}{4} \phi \, C_d(\mathcal{R}) \rho_f \frac{\left|\mathbf{u}^r\right| \, {u}_z^r}{d},\nonumber\\
\rho_f \frac{d  (1-\phi) \left(u_z^f\right)^2}{d  z}&=&(1-\phi) \frac{d  \sigma_{zz}^f}{d  z} - \frac{3}{4} \phi \, C_d(\mathcal{R}) \rho_f \frac{\left|\mathbf{u}^r\right| \, {u}_z^r}{d}. \nonumber
\end{eqnarray}
One can eliminate the fluid normal stress $ \sigma_{zz}^f$ between these two equations and, under the quasi-parallel flow assumption, neglect the left-hand side inertial terms, leading to
\begin{equation}
\frac{d  \sigma_{zz}^p}{d  z} =(\rho_p-\rho_f) g \phi -  \frac{3 \phi C_d(\mathcal{R})}{4(1-\phi)} \; \rho_f \frac{\left|\mathbf{u}^r\right| \, {u}_z^r}{d}
\end{equation}
In the dense regime, this equation reduces to the hydrostatic equation. In the dilute regime, the particle normal stress vanishes and the vertical velocity tends to the settling velocity (see Eq.~\ref{eq_uzr} below).

The horizontal equations of motion read:
\begin{eqnarray}
\rho_p \frac{d  \phi  u_z^p u_x^p}{d  z} &=&\frac{d  \sigma_{xz}^p}{d  z} +\phi \frac{d  \sigma_{xz}^f}{d z} + \frac{3}{4} \phi C_d(\mathcal{R}) \rho_f \frac{\left|\mathbf{u}^r\right| \, {u}_x^r}{d},\nonumber\\
\rho_f  \frac{d  (1-\phi) u_z^f u_x^f}{d  z}&=&(1-\phi) \frac{d  \sigma_{xz}^f}{d  z} - \frac{3}{4} \phi C_d(\mathcal{R}) \rho_f \frac{\left|\mathbf{u}^r\right| \, {u}_x^r}{d}. \nonumber
\end{eqnarray}
Under the quasi-parallel flow assumption, inertial terms on the left-hand side can again be neglected and these equations lead to:
\begin{equation}
\frac{d  \sigma_{xz}^f}{d  z}=-\frac{d  \sigma_{xz}^p}{d  z}=  \frac{3 \phi}{4(1-\phi) } C_d(\mathcal{R}) \rho_f \frac{\left|\mathbf{u}^r\right| \, {u}_x^r}{d}. \label{eq_dsxzfdz} 
\end{equation}
Due to an overall conservation of the momentum flux, the total stress $\sigma_{xz}^p+\sigma_{xz}^f$ is a constant, denoted $\tau$. It can be identified with the fluid shear stress far from the bed, where the particle-borne shear stress vanishes. We also introduce the shear velocity $u_*$, related to the total stress by
\begin{equation}
\tau\equiv\rho_f u_*^2=\sigma_{xz}^p+\sigma_{xz}^f
\end{equation}
The fraction of this momentum flux transported by the diffusive motion of particles is (see Appendix) $(\rho_f u_x^f-\rho_p u_x^p) j_z=\phi (1-\phi) (\rho_f u_x^f-\rho_p u_x^p) u_z^r$. In the limit where the vertical velocity is much smaller than horizontal ones, this effect can be neglected.

Under the quasi-parallel flow assumption, the strain rate modulus is simply the shear rate:
\begin{equation}
|\dot \gamma^f| = {\dot \gamma}^f_{\rm xz}=\frac{d u_x^f}{d z}\quad {\rm and}\quad |\dot \gamma^p| = {\dot \gamma}^p_{\rm xz}=\frac{d u_x^p}{d z}\,.
\end{equation}
The fluid and particle constitutive relations are then simply written as
\begin{eqnarray}
\sigma^{f}_{xz}&=&\rho_f\left(\nu+\ell^2 \left|\frac{d u_x^f}{d z}\right|\right) \frac{d u_x^f}{d z}\label{eq_duxfdz}\\
\sigma^{p}_{xz}&=&\psi(\phi)  \left( \rho_f  \nu+\alpha \rho_p d^2 \left|\frac{d u_x^p}{d z}\right|\right)\frac{d u_x^p}{d z}\label{eq_duxpdz}
\end{eqnarray}
which must be complemented, when the granular phase is unjammed, by the frictional relation  $|\sigma^{p}_{xz}|=\mu(\phi) \sigma^{p}_{zz}$. Combining the vertical and horizontal equations of motion, one obtains an equation relating the derivative of the volume fraction to the vertical velocity:
\begin{equation}
-(\rho_p-\rho_f) g+ \frac{3\;C_d(\mathcal{R})}{4(1-\phi)} \; \rho_f \frac{\left|\mathbf{u}^r \right| }{d} \,  \left(\frac{{u}_x^r}{\mu(\phi)}+{u}_z^r \right)+\frac{\mu'(\phi)\sigma^{p}_{xz}}{\phi \mu(\phi)^2}\; \frac{d\phi}{dz}=0.
\label{eq_phidense}
\end{equation}
Combining it with Eq.~(\ref{eq_phi}), which involves the same quantities, one can deduce ${u}_z^r$ and $d\phi/dz$

\begin{figure}
\centerline{\includegraphics{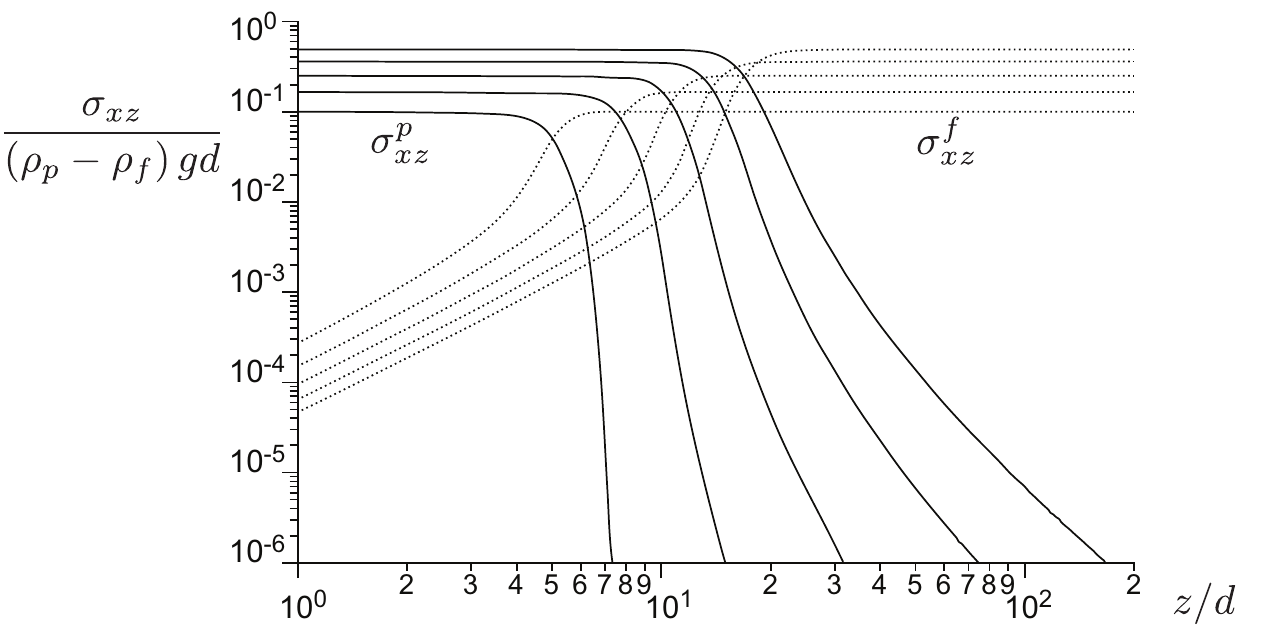}}
\caption{Vertical profiles of the particle-borne shear stress $\sigma_{xz}^p$ (solid lines) and of the fluid-borne shear stress  $\sigma_{xz}^f$ (dotted lines) for the same shear velocity values as in Figure~\ref{Fig2}. The sum of $\sigma_{xz}^p$  and $\sigma_{xz}^f$ is constant, equal to $\rho_f u_*^2$. Note the evolution of the altitude at which $\sigma_{xz}^f= \sigma_{xz}^p$ when $u_*$ is varied.}
\label{Fig3}
\end{figure}

\subsection{Static bed}
The description of the static bed poses a specific problem associated with interparticle friction: it can be prepared at different values of the volume fraction $\phi$, and with different microstructures. This property is at the origin of the hysteresis between the static and dynamic friction coefficients. The problem comes from the formulation of the two-phase flow model, which does not describe the static granular phase.  However, if the static bed is prepared in the critical state, i.e. by shearing the grains at a rate $\dot \gamma$ going to $0$, the volume fraction is equal to the critical value $\phi_c$. We will only consider this situation here. A simple extension would be to consider the bed volume fraction as a parameter, and to impose the continuity of $\phi$ at the static-mobile interface.

We define $z=0$ as the position at which the Coulomb criterion is reached, so that, for $z<0$, the bed is strictly static i.e. the grain velocity is exactly null ($u_x^p=0$ and $u_z^p=0$). Right at $z=0$, the volume fraction reaches the critical volume fraction $\phi_c$. We therefore obtain for $z<0$:
\begin{equation}
\phi=\phi_c \quad {\rm and} \quad P^p =P_0-(\rho_p-\rho_f) \phi_c {g}z.
\end{equation}
The equations describing the fluid in the bed then simplify to
\begin{eqnarray}
\label{eq_statsxz}
\frac{d \sigma_{xz}^f}{dz} &=& \frac{3\phi_c}{4(1-\phi_c)}\;C_d(\mathcal{R}) \rho_f \frac{ \left|{u}_x^f\right|   \,{u}_x^f}{d} \, ,\\
\sigma^{f}_{xz} &=& \rho_f \left(\nu+\ell^2 \left|\frac{d u_x^f}{d z}\right|\right)\; \frac{d u_x^f}{d z}.
\end{eqnarray}

Deep inside the static bed, the fluid velocity  $u^f$ tends to $0$, and one obtains the asymptotic solution in the limit $z\rightarrow-\infty$:
\begin{equation}
\label{eq_sigmafxz}
u_x^f \approx U\,\exp(z/\zeta)\quad{\rm with}\quad  \zeta= \sqrt{\frac{4(1-\phi_c)}{3\phi_c}}\; \frac{d}{s}\;.
\end{equation}
where the coefficient $s$ is defined by Eq.~\ref{dragcoeff}. The flow velocity thus decays over a distance that is a fraction of the grain diameter. $U$ is a shooting parameter that is determined by the matching with the velocity far above the bed. In the same limit, the mixing length equation can be approximated by:
\begin{equation}
\label{eq_ell_asym}
\frac{d \sqrt{\ell}}{dz}\approx \frac{\kappa}{2}\,\sqrt{\frac{u^f}{R_t\nu}} \approx \frac{\kappa}{2}\,\sqrt{\frac{U}{R_t\nu}} \,\exp\left(\frac{z}{2\zeta}\right)
\end{equation}
As $\ell$ must vanish when $z\rightarrow-\infty$, we obtain the asymptotic expression:
\begin{equation}
\label{ell_asym}
\ell \approx \frac{\kappa^2\zeta^2 U}{R_t\nu}\, \,\exp\left(\frac{z}{\zeta}\right)\approx \frac{\kappa^2\zeta^2 }{R_t\nu}\;u_x^f
\end{equation}
%

\begin{figure}
\centerline{\includegraphics{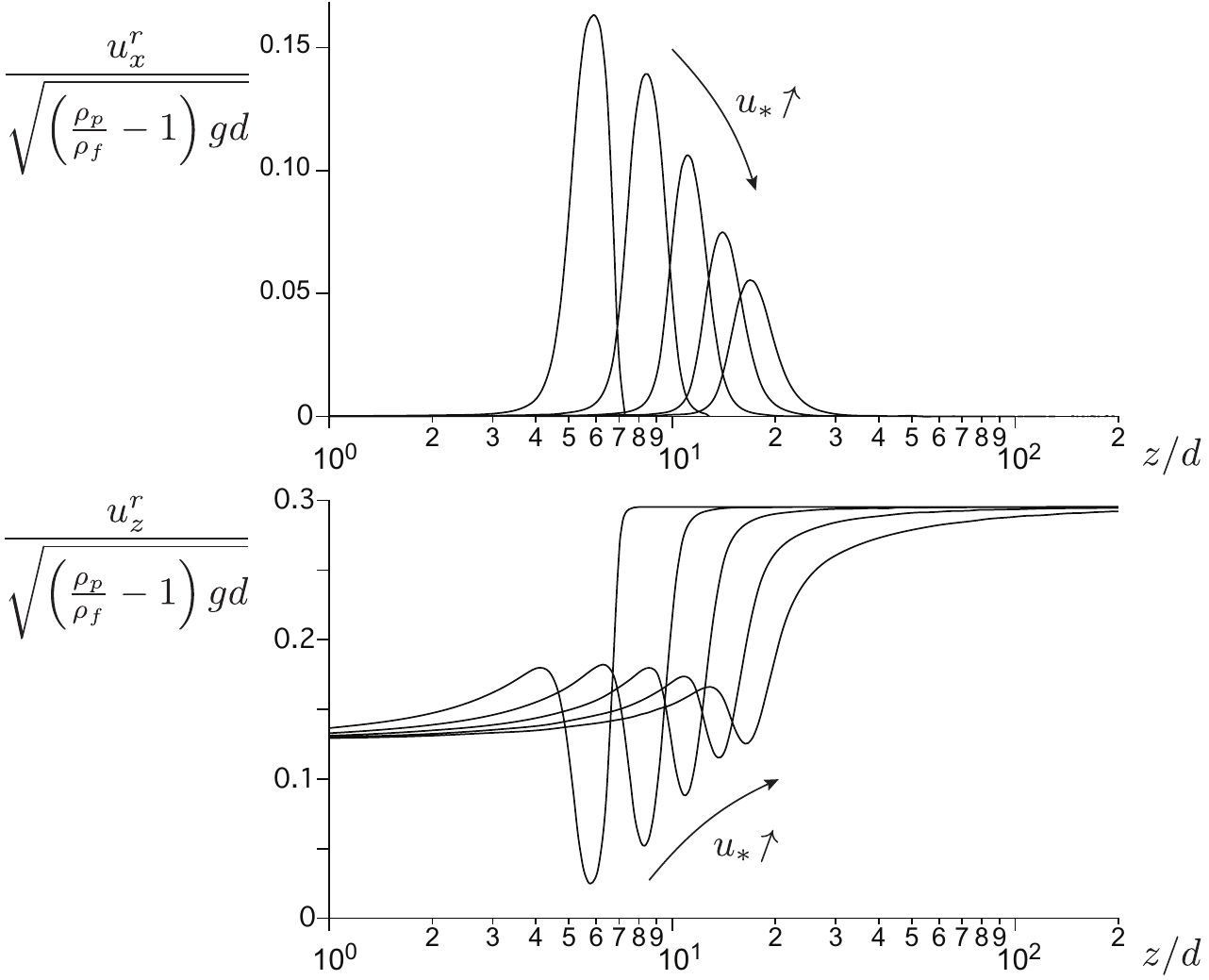}}
\caption{Vertical profiles of the two components of the velocity difference between the two phases. Top: horizontal component $u_x^r$. Bottom: vertical component $u_z^r$. The different lines correspond to the same shear velocity values as in Figure~\ref{Fig2}.}
\label{Fig4}
\end{figure}

\subsection{Dilute zone}
When $z$ tends to $+\infty$, the volume fraction $\phi$, the horizontal component of the relative velocity ${u}_x^r$ and the grain-borne shear stress $\sigma^{p}_{xz}$ all tend to zero. As a consequence, the vertical velocity  $u^r_z=-u^p_z=V_{\rm fall}$ is the settling velocity defined by
\begin{equation}
 \frac{3}{4} \;C_d(\mathcal{R})\; \rho_f \frac{V_{\rm fall}^2}{d} =(\rho_p-\rho_f) g\label{eq_uzr}
\end{equation}
The vertical fluid velocity vanishes like $u^f_z=\phi u^r_z$. The mixing length tends to $ \ell=\kappa (z+z_0)$, where $z_0$, called the hydrodynamic roughness, is a result of the integration. The horizontal velocities tend to:
\begin{equation}
u_x^f \simeq u_x^p \sim \frac{u_*}{\kappa}\ln(1+z/z_0)
\end{equation}
Using the asymptotic expression of the diffusivity $D\sim\kappa u_* (z+z_0)/{\rm Sc}$ and the velocity $u^r_z \sim V_{\rm fall}$, the continuity equation integrates into the well-known volume fraction profile:
\begin{equation}
\phi \sim \frac{\phi_{\rm 0}}{\left(1+z/z_0\right)^{\beta}} \quad {\rm with} \quad \beta=\frac{{\rm Sc} \, V_{\rm fall}}{\kappa \, u_*},
\label{phipowerlaw}
\end{equation}
where the exponent $\beta$ is known as the Rouse number. This asymptotic behaviour selects the only physical solution, and therefore the value of $U$. By contrast, unphysical solutions present a singularity ($\phi \to \phi_c$ or $\phi \to 0$) at finite height. Just like $z_0$, the multiplicative constant $\phi_{\rm 0}$ is selected by the asymptotic expressions derived inside the static bed.
%

\begin{figure}
\centerline{\includegraphics{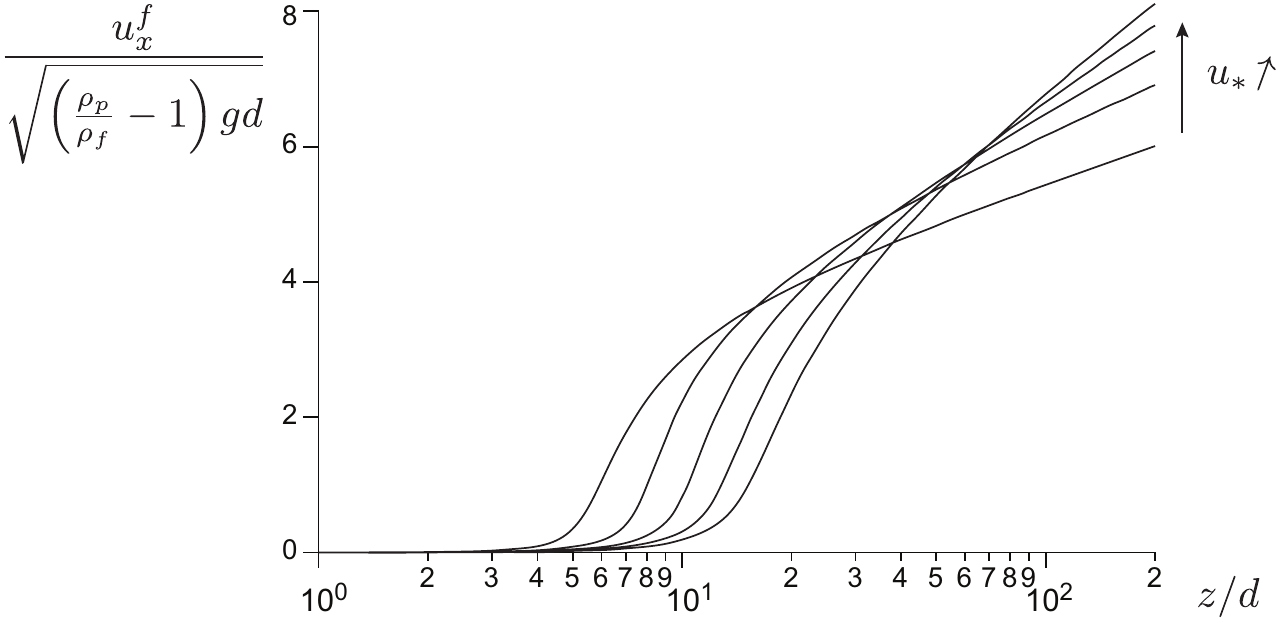}}
\caption{Vertical profiles of the horizontal fluid velocity $u_x^f$, for the same shear velocities as in Figure~\ref{Fig2}.}
\label{Fig5}
\end{figure}

\section{Results}
The  equations governing the evolution of  ${u}_x^f$, $\sigma_{xz}^f$, $\ell$, $u_x^p$ and $\phi$ have been integrated numerically, using a shooting method to satisfy the asymptotic expansions on both sides of the integration domain. In practice, the equations are integrated upward using the Runge-Kutta algorithm. The integration is started from a point deep enough in the static bed for the asymptotic expressions to be valid within numerical errors. The unique solution whose volume fraction decays algebraically at infinity is obtained by bracketing the value of $U$.

\begin{figure}
\centerline{\includegraphics{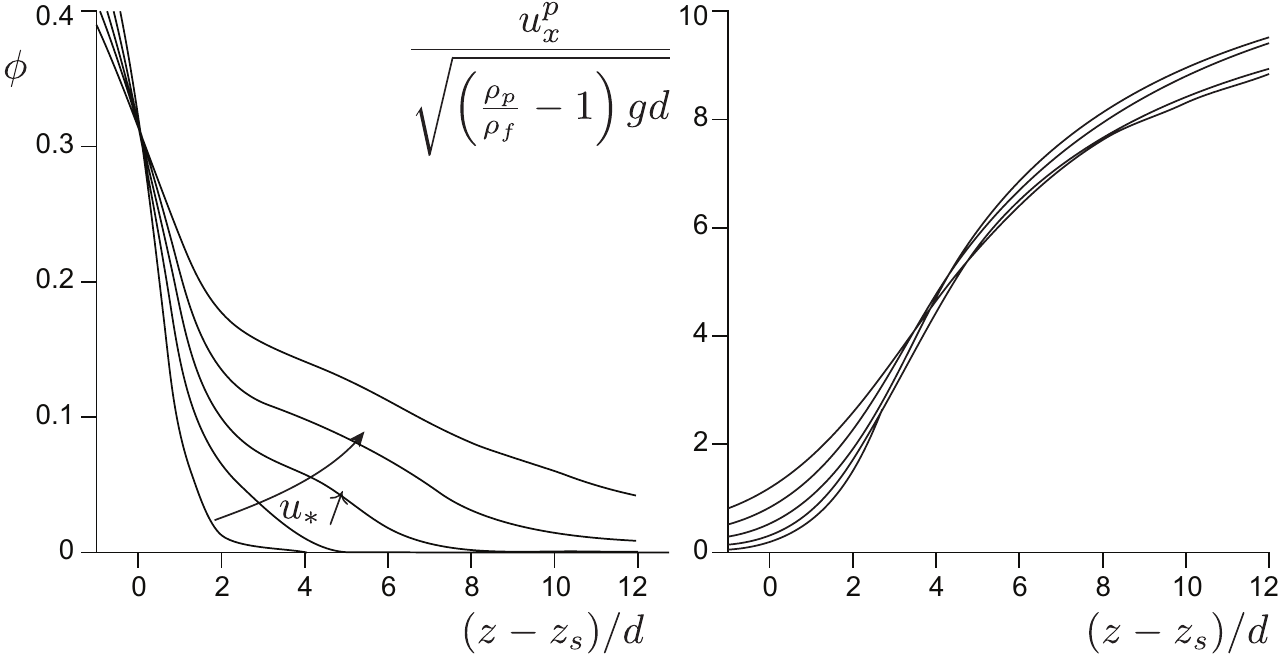}}
\caption{Vertical profiles of the volume fraction (left) and of the horizontal particle velocity $u_x^p$ (right), for the same shear velocities as in Figure~\ref{Fig2}. The position $z$ is measured with respect to the altitude $z_s$ at which $\phi$ crosses $\phi_c/2$.}
\label{FigSuppl}
\end{figure}

We detail and comment on in this section the results obtained for a rescaled viscosity $\nu/\sqrt{\left(\frac{\rho_p}{\rho_f}-1\right) g d^3}=10^{-1}$, or $d/d_\nu \simeq 4.64$. For quartz grains in water, this value corresponds to a mean diameter of $d\simeq 180{\rm \mu m}$. For the shear velocities we have considered (see caption of Fig.~\ref{Fig2}), the viscous length $\nu/u_*$ is a fraction of the grain diameter, in the range $0.14$--$0.33 d$. As shown below, this is small in comparison to the size of the transport layer (typically $5$--$20 d$), so that the curves we present here correspond to transport in the turbulent regime.

By systematically varying the shear velocity or, equivalently, the Shields number, we find that the asymptotic conditions on both sides cannot be matched (i.e. there is no solution) below a threshold value $u_{\rm th}\simeq 0.2\,\sqrt{\left(\frac{\rho_p}{\rho_f}-1\right)gd}$. Experimentally, a similar dimensionless threshold velocity is found, although slightly larger, equal to $0.26$ (see Andreotti et al. 2013 and references therein).

Figure~\ref{Fig2} shows the volume fraction vertical profiles $\phi(z)$ in log-log and lin-lin scales. Close to the static sand bed, grains are transported in a rather dense layer whose thickness increases with $u_*$. This corresponds to a form of bed load where several granular layers are entrained (sheet flow). It is worth noting that the scale separation between the transport layer thickness and the grain diameter is never good enough to expect a quantitive description by a continuum model. Far from the static bed, the volume fraction decreases as a power law of the height, as expected (Eq.~\ref{phipowerlaw}).

Figure~\ref{Fig3} shows the vertical profile of the grain-borne and the particle-borne shear stresses for five values of $u_*$. The sum of $\sigma_{xz}^p$ and $\sigma_{xz}^f$ is constant, equal to $\rho_f u_*^2$. Far above the static bed the whole shear stress is carried by the fluid. Furthermore, close to the static bed, it is carried by the grains and transmitted through contact forces. The transfer of momentum flux from the fluid to the particles occurs at a position that can be interpreted as the top of the bedload layer. One observes in Figure~\ref{Fig4}a) a corresponding peak in the profile of the horizontal velocity difference $u_x^r$, whose position, on the order of $10d$, increases with $u_*$. This relative velocity between the two phases is associated with a horizontal drag force: the momentum flux serves to balance friction and entrain the particles from the bed into motion. Figure~\ref{Fig4}b) shows the vertical velocity profile. As expected, it tends to the settling velocity $V_{\rm fall}$ far above the bedload layer. Inside the bedload layer, one observes that $u_z^r$ is still large, which means that there is a balance between pressure-induced migration and shear-induced diffusion.

We emphasize again that the equations solved are valid throughout the system. The two-layer structure of the transport layer comes out of the solution but was not imposed as in Revil-Baudard \& Chauchat (2012). Also, it is not clear whether the specific sublayers characterized by Berzi (2011, 2013) can be identified here. In particular, $\phi$ is not constant in the bedload layer.

\begin{figure}
\centerline{\includegraphics{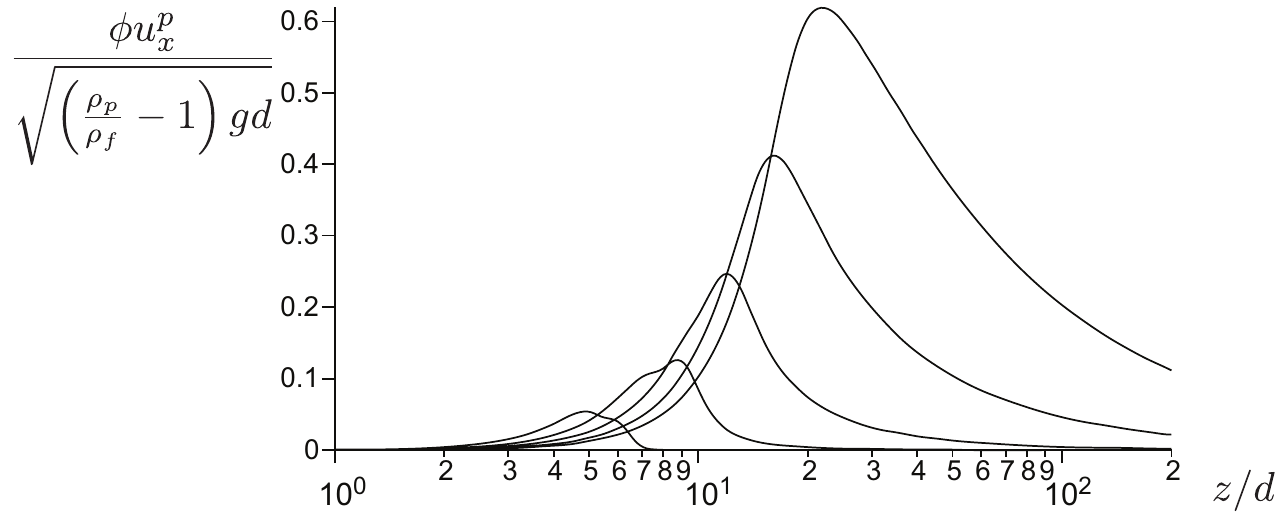}}
\caption{Vertical profiles of the flux density $\phi u_x^p$, for the same shear velocities as in Figure~\ref{Fig2}.}
\label{Fig6}
\end{figure}

The fluid velocity profile $u_x^f(z)$ (see Figure~\ref{Fig5}) is, as expected, logarithmic far above the static bed, but is strongly reduced in the dense transport layer. When particles are accelerated by the flow, the flow, in turn, is decelerated. This negative feedback mechanism takes place over a thicker and thicker region, as $u_*$ increases, which coincides with the bedload layer (Fig.~\ref{FigSuppl}). Because of this strong feedback, the fluid contribution to particle diffusion is negligible close to the static bed. As a consequence, the value of $u_z^r$ when $z \to 0$ is found to be independent of the shear velocity (Fig.~\ref{Fig4} bottom). The structure predicted for the bed load layer resembles qualitatively that observed experimentally by Capart \& Fraccarollo (2011): the volume fraction $\phi$ is not constant but decreases with the height (roughly linearly) while the particle velocity $u_x^p$ increases.

The average sediment transport at the altitude $z$ is characterized by the product of the volume fraction $\phi$ and the particle velocity $u_x^p$. More precisely, $\rho_p \phi u_x^p$ is the density of the mass flux; $\phi u_x^p/\phi_c$ is the density of the volume flux. Figure~\ref{Fig6} shows that $\phi  u_x^p$ presents a peak at the top of the bed load layer, followed by a tail associated with the dilute turbulent suspension. The total sediment flux is obtained by integration over the vertical direction:
\begin{equation}
q=\int_0^\infty \phi \, u_x^p\,dz
\end{equation}
By definition it quantifies the volume of grain crossing a unit length transverse to the flow per unit time. The mass flux is $\rho_p q$ and the volumetric flux at the bed volume fraction is $q/\phi_c$.

\begin{figure}
\centerline{\includegraphics{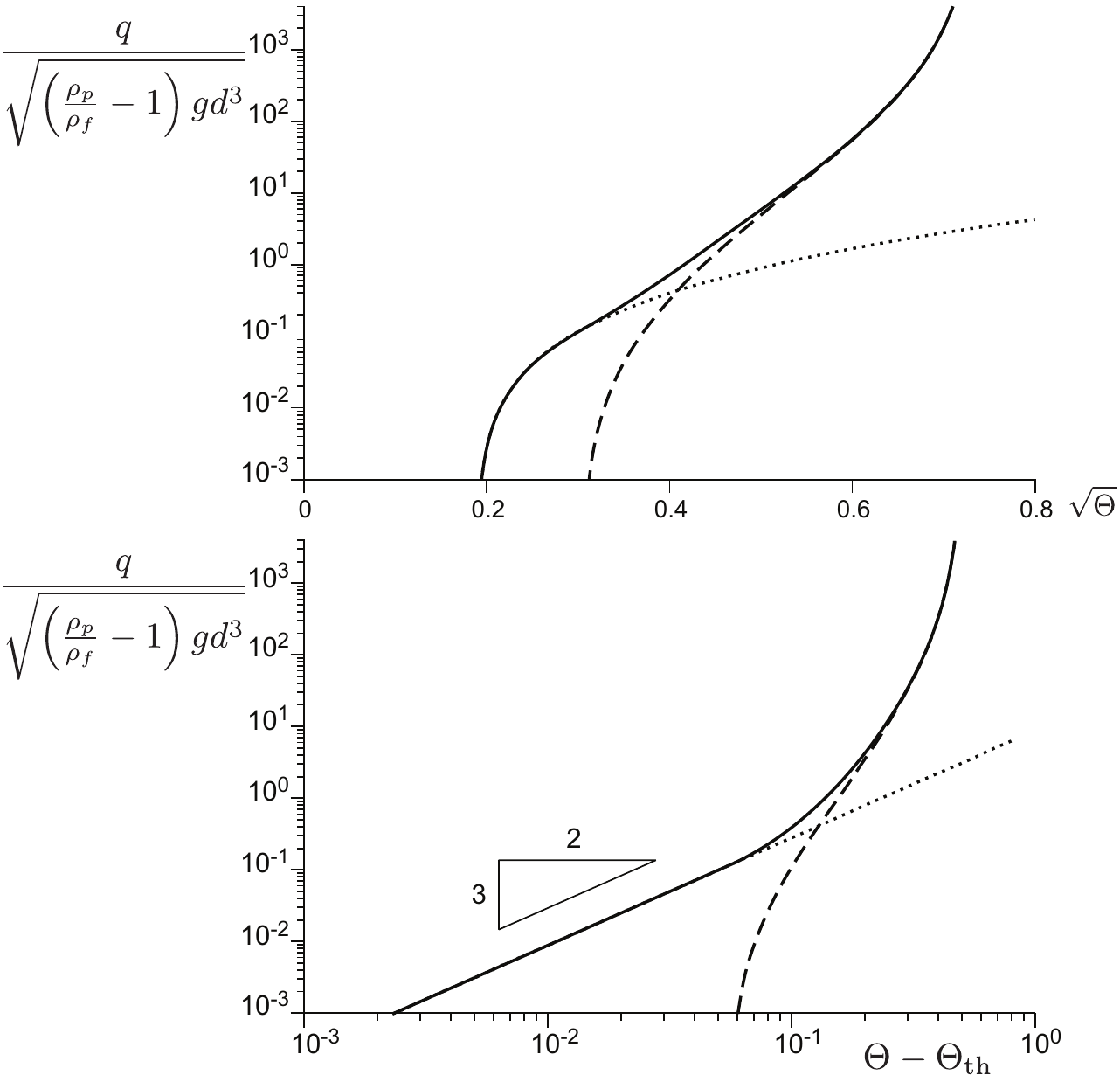}}
\caption{Dependence of the sediment flux $q$ on the Shields number $\Theta$. (a) $q$ as a function of the rescaled shear velocity $\sqrt{\Theta}$. (b) $q$ as a function of $\Theta-\Theta_{\rm th}$. Solid line: total flux. Dashed line: contribution of the suspended load, determined from the asymptotic expansions obtained in the dilute limit (Eq.~\ref{fluxsuspended}). Dotted line: contribution of bed load, determined from calculations performed in the limit $Sc \to \infty$. The bed load flux $q$ follows the Meyer Peter-M\"uller scaling in $(\Theta-\Theta_{\rm th})^{3/2}$.}
\label{Fig7}
\end{figure}

Figure~\ref{Fig7}a) shows the dependence of the sediment flux $q$ on the Shields number $\Theta$. It vanishes below the threshold Shields number $\Theta_{\rm th}$ and beyond, it increases with $\Theta$. It diverges at the Shields number $\Theta_m$ for which the Rouse number $\beta$ reaches $1$. In real conditions, the total sediment flux does not diverge at $\Theta_m$ because it is limited by the finite flow thickness. We have kept here an unbounded flow in order to highlight the transition to a turbulent suspension. Solutions at higher Shields numbers can be deduced by asymptotic matching, when there is a separation of length scale between the flow thickness and the bedload layer thickness.

By integrating the asymptotic expressions obtained far from the sediment bed, we can deduce the contribution of turbulent suspension to the sediment flux:
\begin{equation}
q_{\rm sus}=\frac{\phi_{\rm 0} u_*}{\kappa} \int_0^\infty \left(1+z/z_0\right)^{-\beta}\ln(1+z/z_0)\,dz= \frac{\phi_{\rm 0} u_* z_0}{\kappa (1-\beta)^2} \, .
\label{fluxsuspended}
\end{equation}
With this definition, $q_{\rm sus}$ corresponds to the sediment flux for which the water flow and the concentration profile would behave everywhere as in the asymptotic limit $z \to \infty$. One observes in Figure~\ref{Fig7}a) that $q_{\rm sus}$ (dashed line) indeed becomes dominant around the inflection point of the curve $q(\Theta)$. Conversely, in order to determine precisely the contribution $q_{\rm bl}$ of the bedload to the total flux, we have integrated the equations in the limit of an infinite turbulent Schmidt number ${\rm Sc}$, i.e. without any diffusion of particles resulting from turbulent fluctuations. Importantly, and although this was not given, we have checked that $q_{\rm bl}+q_{\rm sus}$ accurately gives $q$. The bed load contribution is dominant at low shear velocity. It is extremely well fitted by a Meyer \& Peter-M\"uller-like relation:
\begin{equation}
q_{\rm bl}\propto (\Theta-\Theta_{\rm th})^{3/2}
\end{equation}
For the parameters chosen in this paper, the best fit gives a threshold shear velocity of $u_{\rm th}\simeq 0.19\,\sqrt{\left(\frac{\rho_p}{\rho_f}-1\right)gd}$ and equivalently a threshold Shields number of $\Theta_{\rm th}\simeq0.037$. We emphasize again that such a continuum model eventually becomes inaccurate close to the transport threshold, as particle-scale processes become dominant.

\begin{figure}
\centerline{\includegraphics{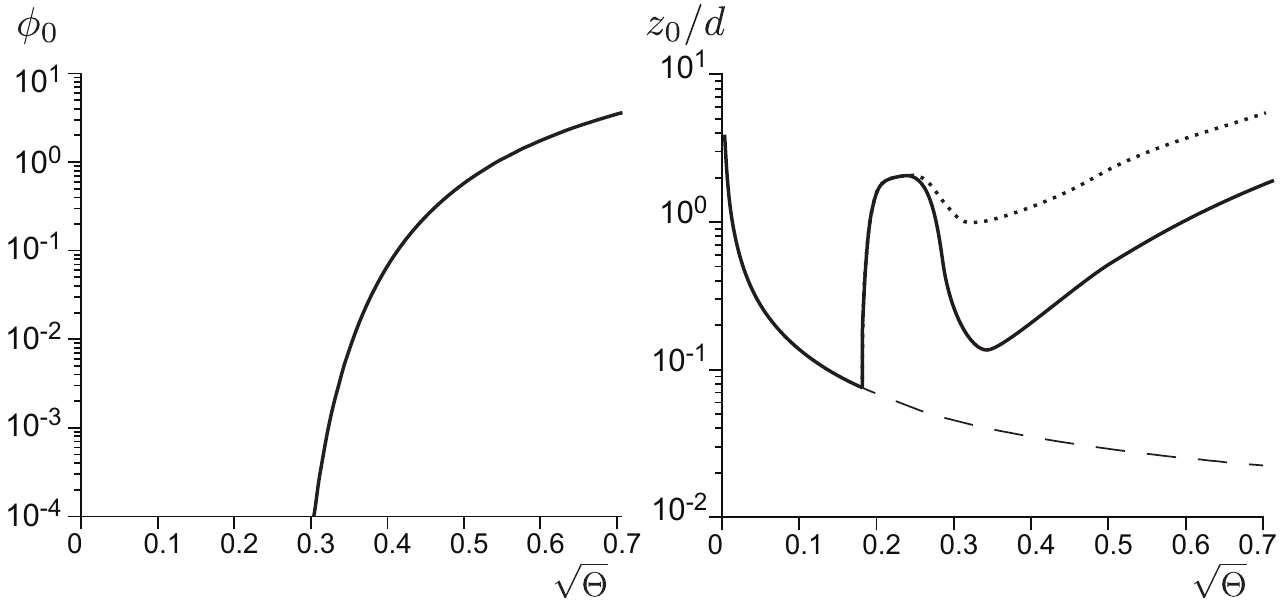}}
\caption{(a) Dependence of the apparent basal volume fraction $\phi_0$ on the Shields number $\Theta$. Starting from the expansion of Eq.~ (\ref{phipowerlaw}) in the region of suspension, $\phi_0$ appears as an extrapolated value of the volume fraction at the bed. From the point of view of the turbulent suspension, $V_{\rm fall}\,\phi_0$ is the effective erosion rate. (b) Dependence of the hydrodynamic roughness on $\Theta$, with (solid line) and without transport (dashed line). Dotted line: roughness obtained with pure bed load, in the limit $Sc \to \infty$. }
\label{Fig8}
\end{figure}

Amongst the novel aspects of our approach, we are able to predict the effective erosion rate seen by the turbulent suspension or, more precisely, the rate at which the bed load particles are injected in this suspension. Figure~\ref{Fig8}a) shows the parameters extracted from the asymptotic expansions in the upper zone. The apparent basal volume fraction $\phi_0$ is defined from Eq.~ (\ref{phipowerlaw}). It increases rapidly and saturates to a value of order $1$ at the suspension threshold. Figure~\ref{Fig8}b) shows the hydrodynamic roughness. In the absence of transport, $z_0$ decreases as $\nu/u_*$. At the transport threshold, it suddenly increases by one order of magnitude before dropping again when suspension becomes dominant. As this quantity can be accessed experimentally at a distance from the bed, it could be used to show the existence of a negative feedback of sediment transport on the flow, in the turbulent regime. Both curves can be prolonged at higher Shields numbers, for a finite flow thickness.

\section{Concluding remarks}
The two-phase flow model derived here is an extension of models previously proposed to describe dense viscous flows and turbulent suspensions. We have added two novelties with respect to these works: the description of dense flows at large Stokes numbers and the  shear-induced diffusion of particles. In the dilute limit, the diffusion is due to hydrodynamic interactions (Leighton \& Acrivos,1987). It had only been qualitatively discussed in the literature (Fall et al., 2010) in the very dense case (for $\phi \simeq \phi_c$). We have shown here that it results from non-affine particle motion and identified its scaling law.

From the point of view of sediment transport, the two-phase flow model derived here is able to describe all regimes: viscous and turbulent, bedload and suspension. Regarding bedload, we have relaxed a hypothesis used in previous models: the volume fraction is not assumed to be homogeneous, equal to $\phi_c$, in the bedload layer. We predict in contrast that the sediment transport layer is rather dilute, with a volume fraction profile continuously decreasing to $0$ away from the static bed. At large shear velocity, when the sediment transport mostly takes place in suspension, our model is able to predict the erosion/deposition rate into the bulk from the equations of mechanics, by matching to the bed load zone. One does not have to introduce a phenomenological erosion rate, balanced by sedimentation, as in previous models of turbulent suspension.

An important future work would be to compare these theoretical predictions with experimental or numerical data. At present, several ingredients of the model involve parameters that have never been measured. For a comprehensive comparison, a first calibration step of these parameters is needed before fitting concentration and velocity profiles (Nnadi \& Wilson, 1992; Sumer et al., 1996; Cowen et al., 2010). For example, the factor $S_\phi$, related to the diffusion of the particles of particles induced by non-affine motion, could be assessed using heterogeneous shear flows (Bonnoit et al., 2010). Also, the predicted erosion/deposition rate, which sensitively depends on the particle shear-induced diffusion, could be tuned using numerical simulations in the spirit of those of Dur\'an et al., (2012).

With this two-phase description, we recover the scaling law phenomenologically proposed by Meyer-Peter \& M\"uller (1948) for bedload transport. However, one expects the model to become inaccurate close to the transport threshold, where the sparse mobile grains hardly constitute an eulerian phase. The analysis of forces and torques on a single grain (Andreotti et al., 2013) seems to provide a better understanding of the transport threshold than the continuum modelling, where the grain size appears indirectly in the decay of the flow velocity inside the static bed. If a fine tuning of parameters can improve the agreement of the model with observations in this range of velocities, it may well be that the dynamics is actually dominated by fluctuations (of the bed surface structure in particular), and can therefore not be described in a mean-field manner.

On the other hand, we expect the two-phase approach to provide an accurate description of the transition from bed load to suspension. It can therefore be applied to different problems of river morphodynamics like the formation of ripples, alternate bars and meanders by linear instability (for recent reviews, see Charru et al (2013) and Andreotti \& Claudin (2013), as well as references therein). Apart from the expression of the saturated flux, as a function of the shear velocity, morphodynamical models must incorporate two important dynamical mechanisms: the dependence of sediment transport on the bed slope and the relaxation of the sediment flux towards saturation. We expect such a two-phase modelling to give some clues to help in the resolution of ongoing controversies on the emergence of bedforms.

\begin{appendix}
\section{Two-phase Reynolds-averaged equations}
We derive here the terms in the two-phase flow description originating from fluctuations. Here, $\phi$, $u_i^p$ and $u_i^f$ are defined as Eulerian averages. The fluctuations (non-affine velocity field and turbulent fluctuations) around these average quantities are denoted $\phi'$, ${u_i^p}'$ and ${u_i^f}'$. The averaging operation is denoted $<.>$. The Reynolds-averaged mass conservation equation gives a diffusive flux:
\begin{equation}
j_i= <\phi' {u_i^p}'>= <\phi' {u_i^f}'>
\label{defj_i}
\end{equation}
The particle momentum equation involves the total stress $\sigma_{ij}^p$ as the sum of the contact stress $s_{ij}^p$ and of the flux of momentum originating from fluctuations,
\begin{equation}
 \sigma_{ij}^p=s_{ij}^p-\rho_p  \left(  \phi <{u_j^p}' {u_i^p}'>+u_i^p j_j+u_j^p j_i+ <\phi'  {u_i^p}' {u_j^p}'>\right).
\end{equation}
The term $-\rho_p \phi <{u_j^p}' {u_i^p}'>$ is the kinetic stress associated with non-affine motion. The two following terms $-\rho_p \phi \left(u_i^p j_j+u_j^p j_i\right)$ quantify the transport of momentum associated with particle diffusion. If the fluctuations are small enough, the last term can be neglected. We work here under this assumption. Similarly, for the fluid, we introduce the viscous stress $s_{ij}^f$ and obtain:
\begin{equation}
 \sigma_{ij}^f=s_{ij}^f-\rho_f  \left((1-\phi) <{u_j^f}' {u_i^f}'>-u_i^f j_j-u_j^f j_i -  <\phi'  {u_i^f}' {u_j^f}'>\right).
 \label{eqappend}
\end{equation}
The term $-\rho_f (1-\phi) <{u_j^f}' {u_i^f}'>$ is the Reynolds stress. The two following terms quantify the transport of momentum associated with particle diffusion. Again the last term can be neglected arguing that the fluctuations remain small.
\end{appendix}

\vspace*{0.3cm}

\noindent
\rule[0.1cm]{3cm}{1pt}

\noindent
This work has benefited from the financial support of the Agence Nationale de la Recherche, grant `Zephyr' ($\#$ERCS07\underline{\ }18).


\end{document}